\documentstyle[12pt]{article}
\begin{document}

\title{\Large\bf Cluster--Exact Approximation of Spin Glass
Groundstates \normalsize}
\author{Alexander K Hartmann
\thanks{e-mail: hartmann\symbol{64}lattice.tphys.uni-heidelberg.de}\\
{\normalsize Institut f\"ur Theoretische Physik }\\
         {\normalsize Ruprecht--Karls--Universit\"at Heidelberg}\\
	 {\normalsize Philosophenweg 19}\\
         {\normalsize 69120 Heidelberg, Germany}}
\date{\today}
\maketitle

\vspace{1cm}
\begin{center}
\end{center}
\vspace{1cm}

\begin{abstract}
We present an algorithm which calculates groundstates of Ising spin
glasses approximately. It works by randomly selecting clusters of
spins which exhibit no frustrations. The spins which were not selected,
contribute to the local fields of the selected spins.  For the
spin--cluster a groundstate is exactly calaculated by using
graphtheoretical methods. The other spins remain unchanged. This
procedure is repeated many times resulting in a state with low energy.
The total time complexity of this scheme is approximately cubic.  We
estimate that the
groundstate energy density of the infinite system for the $\pm J$ model is
$-1.400 \pm 0.005$ (2d) and $-1.766 \pm 0.002$ (3d).
The distribution of overlaps for selected systems is
calculated in order to characterize the algorithm.
\end{abstract}


The combination of frustration and randomness makes it difficult to
find groundstates of spin glasses \cite{review} using numerical
simulations.  In the past years many methods \cite{review95} have been
proposed including the multicanonical ensemble \cite{berg92}, genetic
algorithms \cite{sutton}, a scheme, which uses storing of spin
configurations \cite{krauth}, and an exact algorithm exhibiting
exponential timecomplexity \cite{bara}.  In this paper we present a
method which combines clustering of spins with exact calculation of
groundstates of polynomial solvable problems in order to calculate
approximately groundstates of Edwards--Anderson (EA) spin glasses.

The Hamiltonion of the EA model is given by

\begin{equation}
H = -\sum_{<ij>} J_{ij} \sigma_i \sigma_j - \sum_i B_i \sigma_i
\end{equation}

where the sum goes over nearest neighbours of spins $\sigma_i = \pm
1$.  The exchange interactions $J_{ij}$ are selected at random
according to a probability distribution. The external field $B_i$ can
be site--dependend.

For our computer experiments we used simple cubic lattices of $N=L^d$
spins ($L=$ linear lenght, $d=$ dimension) with periodic boundary
conditions in no external field ($B_i=0$). The interactions are
$J_{ij} = \pm 1$ with equal probability using the constraint
$\sum_{<ij>}J_{ij}=0$.


The algorithm for the approximation of groundstates in frustrated
Ising systems works by taking a spin configuration and calculating
another one, which has a lower or equal energy. This procedure is iterated
many times.

The idea of the scheme for lowering the energy is to choose a cluster
of spins, which exhibits no frustrations. The interaction of the
cluster with the spins at its boundary is included into the local
fields of the cluster--spins.  The groundstate of the cluster is
exactly calculated by using concepts of graph theory \cite{claibo,
knoedel, tarjan}: An equivalent network is constructed \cite{picard},
the maximum flow is calculated with the Ford--Fulkerson algorithm
\cite{ford} and a minimum cut is constructed \cite{hart}. The new
configuration consists of the unchanged spins, which are not in the
cluster, and the groundstate of the cluster. By definition this
procedure cannot increase the energy.

For describing the spin--cluster the Hamiltonian can be rewritten in
the form

\begin{equation}
H_c = -\sum_{<ij>} J_{ij}t_i t_j \sigma^c_i \sigma^c_j -\sum_i B^c_i
t_i
\sigma^c_i + C
\end{equation}

The values of $t_i = 0, \pm 1$ describe the cluster and are used to
handle antiferromagnetic interactions. If spin $\sigma_i=:t_i
\sigma^c_i$ does not belong to the cluster: $t_i = 0$. If $t_i \neq 0$
and $t_j \neq 0$ their signs have to be choosen so that
$J^c_{ij}:=J_{ij}t_i t_j >0$, because only Hamiltonians with positive
exchange interactions can be transformed into an equivalent network.
The local fields $B^c_i$ include the external field $B_i$ and the
interactions with the neighbours of spin $\sigma_i$, which are not in
the cluster:

\begin{equation}
B^c_i = B_i + \sum_{<j>}J_{ij} (1-|t_j|) \sigma_j
\end{equation}

The constant $C$ summarizes the interactions between the
non--cluster spins and can be dropped.  The construction of the
unfrustrated cluster works in the following way: a spin is randomly selected
as seed.  Iteratively neighbouring spins are added, if no
frustration occours. If a cluster cannot be extended, one more
seed--spin is selected. The following algorithm contains the details.
The local variables $\delta_i$ are used to indicate the spins, which
are already detected as neighbours of the cluster.  The set $A$
contains the neighbours of spin $\sigma_i$, which are already in a
cluster, whereas $B$ contains the neighbours, which are added to the
boundary of the cluster.

\begin{table}[h]

\parbox[b]{12.0cm}{
{\bf algorithm} create\_cluster($\{J_{ij}\}$)\\ {\bf begin} }

\hspace*{1.0cm}
\parbox[b]{11.0cm}{
{\bf for all} i {\bf do}\\ {\bf begin}\\
\hspace*{1.0cm}
\parbox[b]{10.0cm}{
initialise $t_i \leftarrow 0$;\\ initialise $\delta_i \leftarrow 0$;}
{\bf end};\\ {\bf while} there are unmarked spins with $\delta_i = 0$
{\bf do}\\ {\bf begin}

\hspace*{1.0cm}
\parbox[b]{10.0cm}{
select one index $i_0$ from the spins with $\delta_i=0$; \\ push $i_0$
on an empty Stack $S$;\\ $\delta_{i_0} \leftarrow 1$;\\ {\bf while}
$S$ is not empty {\bf do} \\ {\bf begin} }

\hspace*{2.0cm}
\parbox[b]{9.0cm}{
	pop one randomly choosen index $i$ from $S$;\\
$A \leftarrow \{j|j$ is neighbour of $i$ and $t_j\neq 0 \}$;\\
{\bf if} $A=\emptyset$ {\bf then} $t_i \leftarrow +1$;\\
{\bf else if} $\forall\, j \in A:\, J_{ij}t_j$ has the same sign $\alpha$
{\bf
then}\\
{\bf begin}\\ 	\hspace*{1.0cm}
\parbox[b]{8.0cm} { 	$t_i \leftarrow \alpha$;\\
 	$B \leftarrow \{j|j$ is neighbour of $i$ and $\delta_j= 0 \}$;\\
 	{\bf for all} $j\in B$ {\bf do}\\
 	{\bf begin}\\ \hspace*{1.0cm}
\parbox[b]{7.0cm}{ 		push $j$ on $S$;\\
$\delta_j \leftarrow 1$;}
{\bf end}; }
 	{\bf end};\\
{\bf else} $t_i \leftarrow 0$; }
\hspace*{1cm}{\bf end}; }

\hspace*{1.0cm}
\parbox[b]{11.0cm}{
{\bf end};\\ {\bf return} ($\{ t_i\}$); }

\parbox[b]{12.0cm}{
{\bf end}; }

\end{table}

A run for calculating a groundstate
consists of choosing randomly an initial configuration or
choosing all spins pointing up, and calculating
new configurations until the energy can not further be lowered. For this in
our experiments
we found as a good criterium that the
energy did not change for the last $n_g$
steps. We used $n_g = N/2$ (2d,3d) as rule of thumb, because in some
very long runs we observed,
that the longest period between two jumps in
energy never exceeded this value and scaled almost linear
with system size.

Because the minimum energy reached in a run depends on the starting
configuration and on the clusters, which were constructed, we performed
several runs for each realization of the random variables $J_{ij}$.
We used that one, which ended on the lowest energy level.
We found 3 runs per realization sufficent, because
by increasing the number of runs, the average
groundstate energy was only lowered about $0.01$ percent.

Because in each step the spin--cluster is randomly built and
by using the algorithm \cite{hart} all of the degenerated
cluster--groundstates have a positive probability to be calculated,
usually in each step a new spin configuration is constructed, even if the
energy remains constant.  So it is possible to explore large areas of
the configuration space.

In figure \ref{e_von_t} the average energy density $e_L(t) = E_L(t)/N$
of 96 realizations is shown as a function of the step number for lattice sizes
$L=4,6,16$ of the 2d system.
One can see, that in the beginning the algorithm approaches very fast
low values. Later the decrease in energy is very small.

It is possible to use other procedures in
selecting the spins of the cluster. We tried three other methods (we call the
first presented algorithm {\em method A}):
\begin{list}{}{\labelwidth2cm \labelsep0.4cm \leftmargin2.7cm
               \listparindent0pt}
\item[method B] Not only neighbours of cluster--spins were added to the stack
$S$, but neighbours of all spins, which were tested, if they can be added
to the cluster.
\item[method C] The spins which were tested,
if they can be added to the cluster
were totaly selected at random from the spins, which were
yet untested.
\item[method D] Methods A and B are applied alternately.
\end{list}

We tested all four methods by calculating groundstates for eight
realizations of $16^2$ systems (3 runs each). In table \ref{tab1}
the average groundstate energy density $e^0$
is displayed. Also we constructed  for the 2d and 3d case
100 respectively 10
spin-clusters for 100 realizations of system sizes $L=4, 6,\ldots,20$.
The average size fractions $n_c/N$ of
the spin--clusters are shown in the last two rows.

\begin{table}[ht]
\begin{center}
\begin{tabular}{r|cccc}\hline
method & A & B & C & D\\
\hline
$e^0$ & 1.425 & 1.425 & 1.415 & 1.427\\
$n_c/N$ (2d) & 0.708 & 0.696 & 0.644 & --\\
$n_c/N$ (3d) & 0.578 & 0.568 & 0.541 & --\\
\hline
\end{tabular}
\caption{Comparison of four methods used to construct spin-clusters}
\label{tab1}
\end{center}
\end{table}

We observed, that for some realizations method A and for others
method B gave lower energies, so it is plausible, that the combination
of both methods gave the best results. So we used method D for all
further experiments.
Maybe better methods for constructing
the cluster would result in a faster convergence and/or lower
energies.


To estimate the groundstate energies of the 2d and 3d $\pm J$ spin
glass we performed calculations for lattice sizes $L=4,\ldots,30$
(2d) and $L=4,\ldots,14$ (3d) with 96 $(4^2,\ldots,20^2,4^3,\ldots, 10^3)$,
respectively 32 $(30^2,12^3, 14^3)$ realizations of the random
variables $J_{ij}$.

In figure \ref{e0_von_l} the groundstate energy density $e^0(L)$ of
the 2d systems is plotted as function of system size. To estimate the
groundstate energy density of the infinite system we performed a
finite-size scaling analysis. A fit to the function $f^0_L(N) =
f^0_{\infty} + c/N$ \cite{berg92}
results in $e^0 = -1.400 \pm0.005$.  This value is
consistent with earlier results from a genetic algorithm $-1.400\pm
0.005$
\cite{sutton}, multicanonical simulations $-1.394 \pm 0.007$
\cite{berg92} pure MC $-1.407 \pm 0.008$ \cite{swend} and
transfer matrix calculations $-1.4024 \pm 0.0012$ \cite{cheng}.

The results for the 3d systems are shown in figure \ref{e0_von_l3d}.
The fit gives a groundstate energy density of $e^0 = -1.766 \pm0.002$.
The results of other authors are $1.7863 \pm0.0028$ \cite{berg92} and
$1.765 \pm 0.01$
\cite{sutton}

As seen in figure \ref{e_von_t}
the time for approaching the groundstate increases rapidly with
the system size.
The average number of steps needed to calculate a groundstate are
shown in figure \ref{t0_von_l} as a function of the number of spins $N$.
For the 2d system (lower curve) a fit to a function
$t(N) = a + b N^c$ results in a timecomplexity of
$O(n^{1.17 \pm 0.06})$.  The result for the 3d system is similiar:
$O(n^{1.27 \pm 0.08})$.
Because the time to calculate one groundstate of the cluster
increases quadraticly with the number of spins \cite{hart} and the
size of the cluster is linear in the system size, this totally results
in an approximate cubic timecomplexity of the algorithm.
Performing one run for example for a $10^3$ system consisting of 1000
configurations needed about 10 hours on a 80 Mhz PowerPC processor
system.

In order to characterize the way the algorithm approaches the
groundstates we selected one realization of a 2d ($L=16$) system and
used configurations of different energy densities $e_s \in
[-1.414,1.0]$ as starting--configurations. For each configuration 8
different runs for constructing groundstates were performed. For each
run, which resulted in the groundstate energy density value of
$e^0=-1.414$ the last 100 configurations were stored. The overlap

\begin{equation}
q:=\sum_{i=1}^N \sigma^{\alpha}_i \sigma^{\beta}_i
\end{equation}

was calculated between all
pairs $(\alpha,\beta)$ of groundstates belonging to a
starting-configuration .  We obtained the corresponding probability
distributions $P_L(q)$ by counting the numbers of overlaps within
intervals of length $\Delta q=0.05$. The result is displayed in figure
\ref{overlap} for the starting energies $e_s = -1.414$, $-1.07$,
$0.117$.  For these three staring energies 800, 300 respectively 400 of the
resulting states exhibited the lowest energy density. One can see, that
the algorithm approaches local minima of the energy landscape. The
higher the starting energy the more minima are reachable. If a system
is caught in a local minimum, only a restricted area of configuration space
can be explored by the algorithm, although in each step a different
configuration is generated which is indicated by $P_L(1)=0$.  But for
other realizations we got also different distributions.  There are
systems where $P_L(q)$ is always narrow or always broad, independent of
the starting energy.  The
accessible area is usually smaller
for lower groundstate energies than for higher lying groundstates.


In conclusion, in this letter we have presented an algorithm for the
approximation of groundstates for Ising spin glasses. Its time
complexity is approximately cubic in the number of spins.  Because by
each step a large number of spins is allowed to flip, the algorithm
approaches low energies very rapidly.

So it should be interesting to try combinations with other algorithms
like genetic algorithms or Monte Carlo simulations, or to use it for
other Ising type optimization problems.

Although we applied it to the $\pm J$ model it is also possible to
treat more general forms of lattices, interactions and their
probability distributions.

It should possible to get results for $B\neq 0$ in a
resonable amount of time in order to characterize better the phase
diagramm along the $T=0$ axis.

\section*{Acknowledgments}

We would like to thank D W Heermann, K D Usadel and G
Reinelt for fruitfull discussions, A Linke for a critical reading of the
manuscript and the Paderborn Center of Parallel
Computing for the allocation of computer time on a PARIX--PowerPC System.

\clearpage

\section*{Figure Captions}

\begin{itemize}
\item[Figure \ref{e_von_t}]
Average energy density $e_L(t) = E_L(t)/N$ of 2d $\pm J$ spin glass
at different iterations $t$. Shown are 3 sizes $L=4,6,16$.

\item[Figure \ref{e0_von_l}]
Average groundstate energy density
$e^0(L) = E^0(L)/N$ of 2d $\pm J$ spin glass as
a function of size $L$.

\item[Figure \ref{e0_von_l3d}]
Average groundstate energy density
$e^0(L) = E^0(L)/N$ of 3d $\pm J$ spin glass as
a function of size $L$.

\item[Figure \ref{t0_von_l}]
Average number of steps needed to reach a groundstate as function of
system size $N$ for
2d system ($L=4,\ldots,20$).

\item[Figure \ref{overlap}]
Distribution of overlaps for one single realization (2d, $L=16$, $e^0=-1.414$).
For different starting configurations with energies
$e_s =$ $0.117$, $-1.07$, $-1.414$
with 8 independent runs 100 groundstates were calculated. The states with
energy density $e^0=-1.414$ were included in the calculation of the
distribution, i.e. 400, 300, 800 states for the three values of $e_s$.
The lines are guides for the eyes only.
\end{itemize}

\clearpage

\section*{Figures}

\setcounter{figure}{0}
\refstepcounter{figure}

\begin{figure}[h]
\setlength{\unitlength}{0.240900pt}
\ifx\plotpoint\undefined\newsavebox{\plotpoint}\fi
\sbox{\plotpoint}{\rule[-0.200pt]{0.400pt}{0.400pt}}%


\label{overlap}
\end{figure}
\clearpage
\refstepcounter{figure}


\clearpage

\begin{thebibliography}{99}

\bibitem{review} For reviews see \\
Binder K and Young A P, Rev. Mod. Phys. {\bf 58}, 801--976 (1986); \\
Fisher K H and  Hertz J A, Spin Glasses, Cambridge University Press (1991)

\bibitem{review95} Rieger H, in Annual Reviews of Computational Physics
{\bf II}, ed. Stauffer D, World Scientific, Singapore (1995)

\bibitem{berg92} Berg B A and Celik T, Phys. Rev. Lett. {\bf 69},
2292--2295 (1992); \\
Berg B A, Celik T and Hansmann U, Europhys. Lett. {\bf 22}, 63--68 (1993); \\
Hansmann U H E and Berg B A, Int. J. Mod. Phys. C {\bf 5}, 263--265
(1994)

\bibitem{sutton} Sutton P, Hunter D L and Jan N, J. Physique I {\bf 4},
1281--1285 (1994)

\bibitem{krauth} Krauth W and Pluchery O, J. Phys. A {\bf 27},
L715--L720 (1994)

\bibitem{bara} Barahona F, Phys. Rev. B {\bf 49}, 12864--12867 (1994)

\bibitem{claibo} Claiborne J D, Mathematical preliminaries
for computer networking, John Wiley \& Sons, New York 1990

\bibitem{knoedel} Kn\"odel W, Graphentheoretische Methoden
und ihre Anwendung, Springer, Berlin 1969

\bibitem{tarjan} Tarjan R E, Data Structures and Network
Algorithms, Society for Industrial and Applied Mathematics,
Philadelphia, 1983

\bibitem{picard} Picard J P and Ratliff H D, Networks {\bf 5},
357--370 (1975)

\bibitem{ford} Ford L R and Fulkerson D R, Canadian J. Math.
{\bf 8}, 399--404 (1956)

\bibitem{hart} Hartmann A K and Usadel K D, Physica A {\bf 214},
141--152 (1995)

\bibitem{swend} Wang J--S and Swendsen R H, Phys. Rev. B {\bf 38},
4840--4844 (1988)

\bibitem{cheng} Cheung H--F and McMillan W L, J. Phys C {\bf 16}, 7027--7032
(1983)
\end{thebibliography}
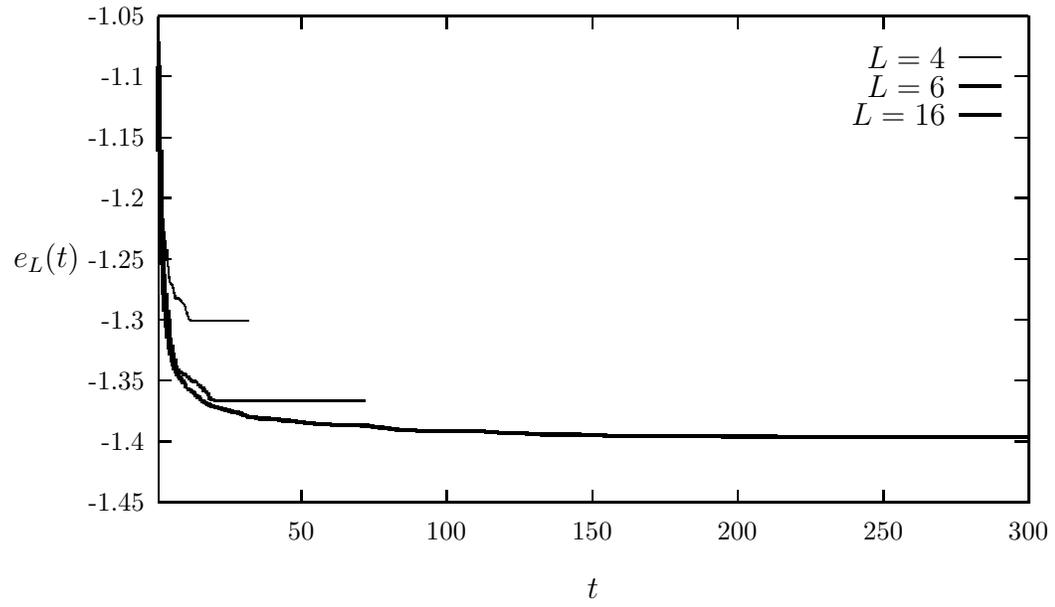
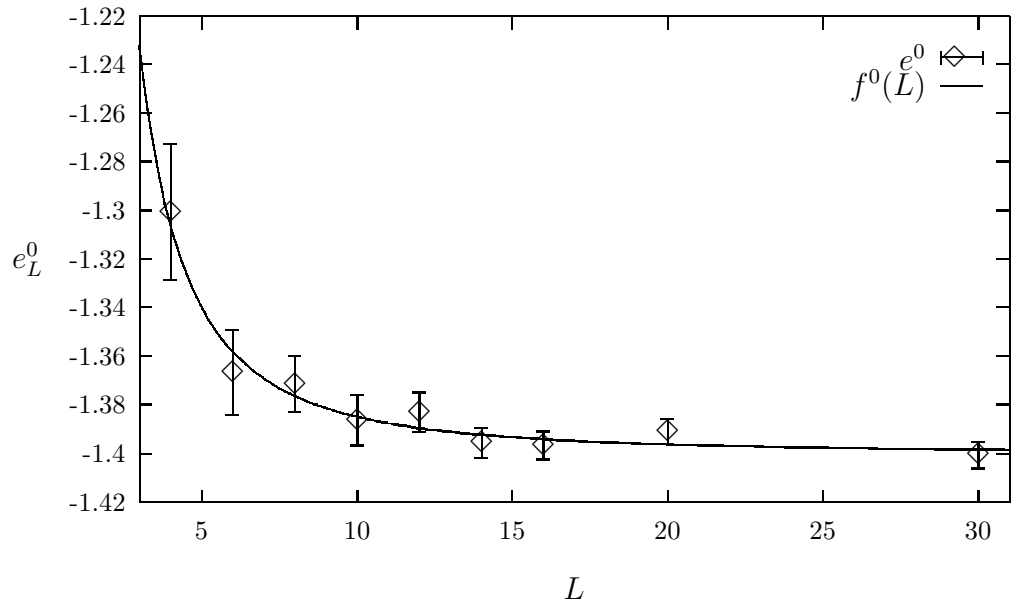
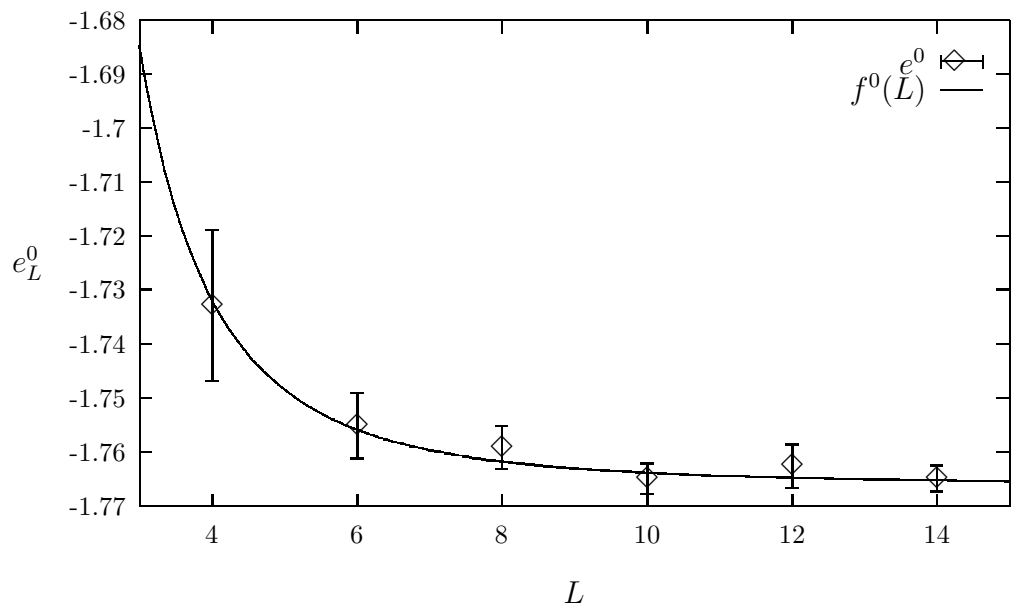
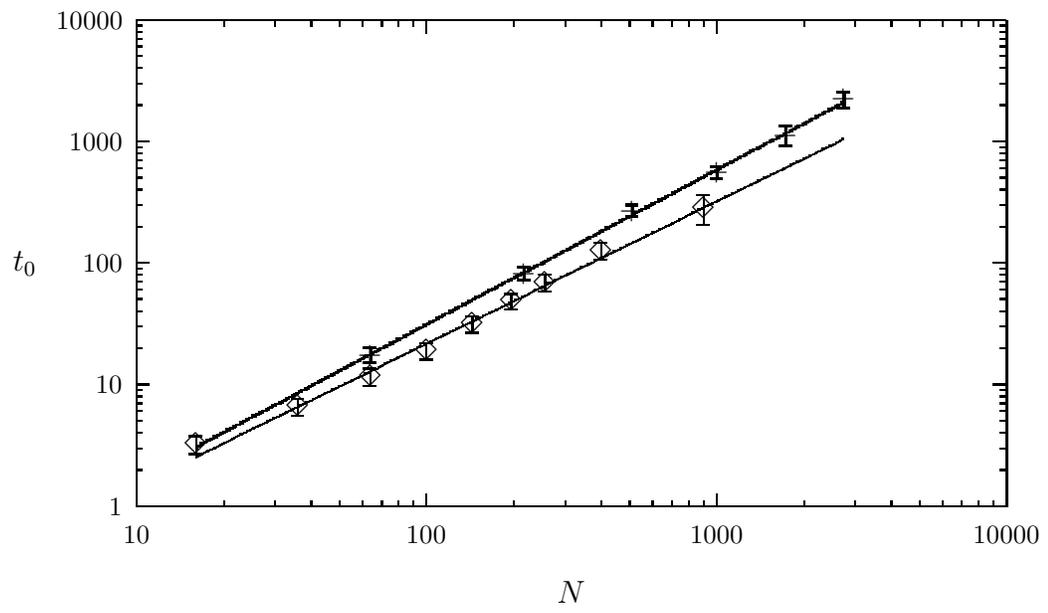
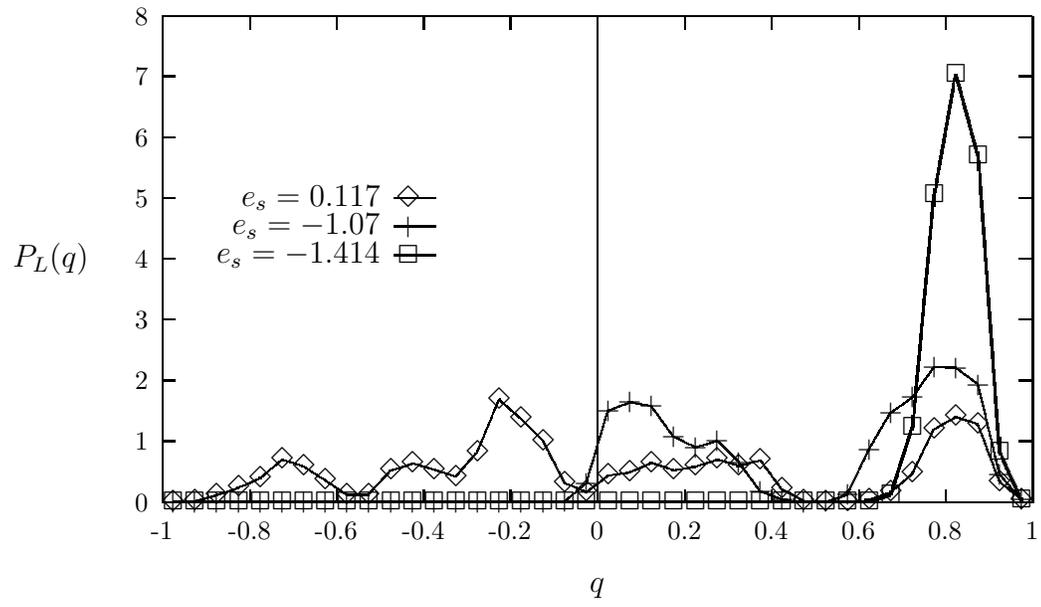
\end{document}